\documentclass{article}
\usepackage{graphicx}
\usepackage{fullpage}

\usepackage{xspace}
\usepackage{url}
\usepackage{xcolor}
\usepackage{subcaption}
\usepackage{comment}

\newcommand{\SM}{Supplementary Material\xspace}

\newcommand{\ignore}[1]{}

\usepackage{listings}
\newcommand{\vir}[1]{``#1''}

\title{FASTA/Q Data Compressors for MapReduce-Hadoop Genomics: Space and Time Savings Made Easy - Version 1}

\author{
Umberto Ferraro Petrillo\thanks{Dipartimento di Scienze Statistiche, Universit\`{a} di Roma - La Sapienza, Rome, 00185, Italy} \hspace{0.05mm} \thanks{To whom correspondence should be addressed.} \and 
Francesco Palini\footnotemark[1] \and 
Giuseppe Cattaneo\thanks{Dipartimento di Informatica, Universit\`{a} di Salerno, Fisciano (SA), 84084, Italy} \hspace{0.05mm} \thanks{Those two authors contributed equally to the research.} \and
Raffaele Giancarlo\thanks{Dipartimento  di Matematica ed Informatica, Universit\`{a} di Palermo, Palermo, 90133, Italy} \hspace{0.05mm} \footnotemark[4]
}

\date{}

\begin{document}

\maketitle

\begin{abstract}
\textbf{Motivation:} Storage of genomic data is  a major cost for the Life Sciences, effectively addressed mostly via  specialized data compression methods.   For the same  reasons of abundance in data production, the use of Big Data technologies is seen as the future for genomic data storage and processing, with MapReduce-Hadoop as leaders. Somewhat surprisingly, none of the  specialized FASTA/Q compressors is available within Hadoop. Indeed, their deployment there  is not exactly immediate. Such a State of the Art is problematic. \\ 
\textbf{Results:} We provide major advances in two different directions. Methodologically, we propose  two  general methods, with the corresponding software, that make very easy to deploy a specialized FASTA/Q compressor within MapReduce-Hadoop for processing files stored on the distributed Hadoop  File System, with very little knowledge of Hadoop. Practically, we provide evidence that the deployment of those specialized compressors within Hadoop, not available so far, results in major cost savings, i.e., on large plant genomes, $30\%$  less HDFS data blocks (one block=128MB), speed-up of at least $x1.5$ in I/O time and  comparable or reduced network communication time  with respect to the use of generic compressors available in Hadoop. Finally, we observe that these results hold also for the Apache Spark framework, when used to process FASTA/Q files stored on the Hadoop File System.\\
\textbf{Contact:} umberto.ferraro@uniroma1.it
\end{abstract}


\section{Introduction}
%
%

Data Compression and the associated techniques coming from Information Theory, has a long and very influential history for the storage and mining of biological data \cite{giancarloCompression09}. In recent years,  it has received increasing attention via the proposal of novel specialized compressors, due to the facts that (a) storage costs have become quite significant given the massive amounts of data produced by HTS technologies (see \cite{Fritz11} for an enlightening analysis, which is still valid \cite{pavlichin2018}); (b) generic compressors, even of the latest generation, e.g. LZ4 \cite{Lz4}, BZIP2 \cite{Bzip2}, are inadequate for  the task of biological data compression.   Good analytic reviews of the State of the Art are provided in \cite{giancarlo2014compressive,Numanagic2016}, although no clear winner compressor has emerged. 
It is worth of mention that data compression is now considered also as something convenient to speed-up the processing of Bioinformatics pipelines. Indeed, following the idea of computing on compressed data, developed in Computer Science under the term Succinct Data Structures \cite{navarro2016}, the concept of Compressive Genomics has been proposed, with some highly specialized proof of principle \cite{loh2012}.

Due to the same reasons of massive data production, Big Data Technologies for Genomics and the Life Sciences,   have been indicated as a direction to be actively pursued \cite{kahn2011future}, with MapReduce \cite{dean2008mapreduce}, Hadoop \cite{HadoopGuide} and Spark \cite{SparkGuide}  being the preferred ones \cite{Cattaneo19}. This is not  just a following of a \vir{Big Data trend} that has proved successful in other fields of Science, since Bioinformatics solutions based on those techniques can be more effective than classic HPC ones, thanks to their  scalability with available hardware \cite{KCH} and to their easiness of use.
For later reference, it is worth pointing out that those technologies have \vir{compression capabilities}
via built-in generic data compressors, e.g., BZIP2 \cite{Bzip2}. The corresponding software is referred to with the technical term Codec, where compression is coding and decompression is decoding. Moreover, although with quite some knowledge of those technologies, it is possible to add other  compressors to  Hadoop, i.e., additional Codecs. It is to be added that not all data compressors are amenable to  a profitable  incorporation due to the requirement of {\em splittable} compression: a file is divided into (un)compressed data blocks that can be compressed and decompressed separately granting in any case the integrity of the entire file. Indeed, processing files compressed using a non-splittable format is still possible under Hadoop, but at a  cost of very long decompression times (data not shown but available upon request). Further discussion on those topics is in Section \ref{sec:split}.

In order to make a compressor splittable,  when its standard version is not, requires  major code reorganization and  rewriting.  In what follows, the term {\em standard} denotes a compressor that executes on a sequential machine, i.e., a PC. 
Given the above discussion about Data Compression, it is rather surprising  that the deployment of specialized compressors for biological data in Big Data technologies  is episodic, in particular for FASTA/Q file formats, e.g., \cite{shi2016}, that host a substantial part of genomic data.

\subsection{Methodological Contributions}
We provide two contributions for the deployment of standard specialised compressors for FASTA/Q files within MapReduce-Hadoop, together with the corresponding software. 

\begin{itemize}
    \item {\bf Splittable Compressor Meta-Codec.} When a standard compressor is splittable, we provide a method that  facilitates  its  incorporation  in Hadoop. Use of  the software library associated to the method  offers  a substantial  savings of programming time for a rather complicated task. Intuitively,  the {\bf  Splittable Compressor Meta-Codec} performs a transformation of  a standard splittable compressor in an Hadoop splittable Codec for that compressor. 
    
\item {\bf Universal Compressor Meta-Codec.} Independently of being splittable or not, as  long as  some mild assumptions in regard to input/output handling, we provide a method to incorporate a data compressor in Hadoop, making it splittable. 
It is worth pointing out that the vast majority of standard specialized FASTA/Q compressors are not splittable. Again, intuitively, the {\bf Universal Compressor Meta-Codec} performs a transformation of a standard compressor in an Hadoop splittable Codec for that compressor. 
\end{itemize}


A few comments are in order.   The 
 {\bf   Splittable Compressor Meta-Codec} provides a template useful for accelerating and simplifying the development of specialized Hadoop Codecs.  The {\bf Universal Compressor Meta-Codec}  allows to support in Hadoop any standard compressor with no programming at all, provided that is usable as a command-line application. The first option has to be preferred when interested in achieving the best performance possible, at a cost of analyzing the internal format employed by files processed with that compressor and writing the required integration code. The second option allows to almost instantaneously support any command-line compressor, but at a cost of possibly reduced performance that we have measured to be negligible with respect  to the direct use of the {\bf Splittable Compressor Meta-Codec}.  
Both methods work also for Spark, when it uses the Hadoop File System. 
Finally, given the pace at which new standard specialized compressors are implemented, our methods  can readily support the deployment of those future implementations in Hadoop.

For later use, we refer to the version of a standard compressor  with the prefix HS  when the incorporation in Hadoop has been made by using the {\bf Splittable Compressor Meta-Codec} or an Hadoop splittable Codec is already available,  e.g., LZ4 becomes HS\_LZ4. Analogously, we use the prefix HU, when  the {\bf Universal Compressor Meta-Codec.} has been used.

.

\subsection{Practical Contributions}
We provide experimental evidence that our methods  are  a major advance in dealing with massive data production in genomics  within one of the Big Data technologies of choice. Indeed, for the {\bf Universal Compressor Meta-Codec},   we show the following via an experimental  comparative analysis involving a selection of specialized FASTA/Q compressors vs the generic compression Codecs already available in Hadoop.

\begin{itemize}

\item{\bf Disk space savings.} The size of the FASTA/Q files  is significantly reduced with the use of specialized HU Codecs  vs the  generic HS  available in Hadoop.  Consequently, the cost of the hardware required to store them in the Hadoop File System  is reduced. 

\item{\bf Reading time savings.} When using a specialized  HU, the additional time required to decompress a FASTA/Q file in memory is counterbalanced by the much smaller amount of time required to load that file  from the Hadoop File System. This  results  in a significant reduction of the overall reading time. 

\item{\bf Network communication time overhead  savings.} The number of concurrent tasks required to process, in a distributed way,  a  FASTA/Q file compressed via an HU is  greatly reduced, thus allowing for a significant reduction of the network communication time overhead required for the recombination of their outputs.

\end{itemize}

As for the 
{\bf Splittable Compressor Meta-Codec}, we reach the same conclusions as above, but the experimentation is somewhat limited: the only standard specialized compressor for FASTA/Q files featuring a splittable format is DSRC \cite{roguski2014dsrc}.  Finally, disk space and reading time savings apply also to the Apache Spark framework, when used to process FASTA/Q files stored on the Hadoop File System.

\section{Methodologies}


This section is organized as follows. Section \ref{sec:split} is dedicated to introduce some basic notions about Hadoop, useful for the presentation of our methods. Section \ref{subsec:guidelines} outlines some technical problems regarding the design of a splittable Codec for Hadoop, proposing our solutions. The last two section are dedicated to the description of our two Meta-Codecs.

\subsection{Preliminary}\label{sec:split}
MapReduce is a programming paradigm for the development of algorithms able to process Big Data on a distributed system on an efficient and scalable way. It is based on the definition of a sequence of {\em map} and {\em reduce} functions that are executed, as {\em tasks}, on the nodes of a distributed system. Data communications between consecutive tasks is automatically handled by the underlying distributed computing framework, including the {\em shuffle} operation, required to move data from one node to another one of the distributed system.

In Section 1 of the \SM we provide more information about this topic, including Hadoop, one of the most popular MapReduce implementation. Here we limit ourselves to describe how files are stored in the the Hadoop File System, i.e. HDFS.

When uploading a large file to HDFS (by default, larger than $128$MB), it is automatically partitioned into several parts of equal size, where each part is called {\em HDFS data block} and is physically assigned to a Datanode, the nodes of the distributed system that execute map and reduce tasks.

For fault-tolerance reasons, HDFS data blocks can be replicated on several Datanodes according to a user-defined {\em replication factor}. This allows to process a HDFS data block even if the Datanode originally containing it becomes unavailable. By default, Hadoop assumes that each map task processes only the content of one particular HDFS data block. However, it may happen that, because of the aforementioned partitioning, a record to be analyzed by one map task is cut into two parts located in two different HDFS data blocks. We refer to these cases as {\em disalignments}. 

This circumstance is managed by HDFS through the introduction of the {\em input split} concept or {\em split}, for short. It can be used, at the application level, to logically redefine the range of data to be processed by each map task, thus allowing a map task to process data found on HDFS data blocks different than the one it is processing.

\subsubsection{Hadoop Support for the Input of Compressed Files}
 Currently, Hadoop supports two types of Codecs:

\begin{itemize} 

\item \emph{Stream-oriented.} Codecs in this class  require that the whole file  be available to each map task prior to decompressing it. For this reason, when a map task starts its execution, a  request is issued to the other nodes of the cluster. As a result, all the parts of the file to be processed are collected from these nodes and merged into a single local file. This type of Codec can be developed by creating a new Java class implementing the standard Hadoop \texttt{CompressionCodec} interface.

\item \emph{Block-oriented.} Codecs in this class  allow each map task to decompress only a portion of the input file, without requiring the remaining parts of it. They assume the compressed file to be logically split into data blocks, here referred to as  {\em compressed data blocks}, where each of them  can be decompressed independently of the others. Assuming the possibility of knowing the boundaries of each compressed data block, a map task can autonomously extract and decompress all the compressed data blocks existing in its HDFS data blocks.  This type of Coded can be developed by creating a new Java class implementing the standard Hadoop \texttt{SplittableCompressionCodec} interface.

It is worth noting that the stream-oriented approach implies a significant computational overhead, as the same file is decompressed as many times as the number of map tasks processing it. It implies also a significant communication overhead, because the same file has to be replicated on each computational node running at least a map task. Finally, it may prevent a job from running at all because map tasks may not have enough memory to handle the decompression of the input file (e.g., when handling large files). For this reason, in this research,  we focus on block-oriented Codecs, i.e.,  \emph{splittable} Codecs.

\end{itemize}

\label{sec:Methods}

\subsection{General Guidelines for the Design of an Hadoop Splittable Codec.}
\label{subsec:guidelines}

Here we consider some problems that a programmer must face in order to obtain an Hadoop splittable Codec, offering solutions. We concentrate on genomic files, although the guidelines apply to any lossless textual compressor. 

There are two problems to face when extracting genomic sequences from a splittable compressed file. The first is about inferring the logical internal organization of the compressed file in regard to determine the relative position of the compressed data blocks. The second is in regard to the management of the possible disalignments existing between the physical partitioning of the file, as determined by HDFS, and the internal logical organization of the compressed file in compressed data blocks. 
 In Section \ref{subsec:inferring} and in Section \ref{subsec:disalignments}, respectively, these problems are described in details and the solution we propose is presented.


\begin{figure}[ht]
    \centering
    \includegraphics[scale=.25]{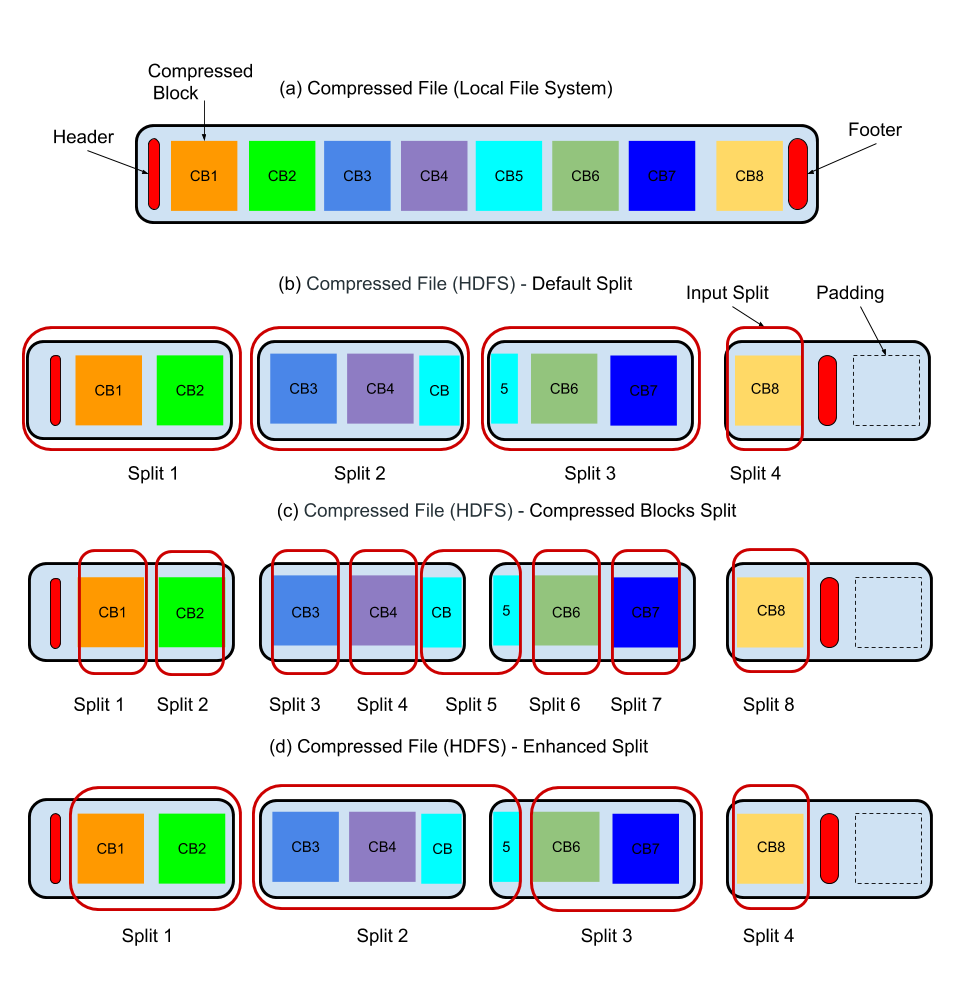}
    \caption{The layout of a block-oriented compressed data file when uploaded to HDFS. In the figure, (a) the original file includes an header, a footer and $8$ compressed data blocks. (b) When uploaded to HDFS, it is partitioned into $4$ HDFS data blocks. (c) As a result of the partitioning, the  compressed data block labeled as $CB{5}$ is divided into two parts and assigned to two different HDFS data blocks. Using the {\em Compressed Block Split} strategy, each compressed data block is modeled as a distinct split. (d) Using the {\em Enhanced Split} strategy, several compressed data blocks are grouped into fewer input splits.}
        \label{fig:layout2}

\end{figure}

\subsubsection{Determining the Internal Structure of a Compressed File}
\label{subsec:inferring}

A map task can extract and decompress the compressed data blocks existing in the HDFS data block it is analyzing only if it knows their size and relative positions. However, this information could be stored elsewhere (e.g., in the footer of the compressed file) or it could be encoded implicitly.  

In the following, we provide a solution for efficiently dealing with the most frequent scenario, i.e., the one where the list of compressed data blocks is made explicitly available. We refer the interested reader to \cite{Bzip2} for an example of a solution for encoding this list implicitly.


\paragraph{\em Explicit Representation.} An explicit list of all the compressed data blocks existing in a compressed file is maintained in an auxiliary {\em index} data structure. This latter may either be located at the beginning or at the end of the file (e.g., DSRC \cite{roguski2014dsrc}), or it can be saved in multiple copies along a file.  In some other cases, this data structure can be saved in an external file complementing the compressed file.



In this case, the solution proposed here is to have one process to retrieve the index before processing the compressed file and send a copy to all nodes of the distributed system using the standard Hadoop {\tt Configuration} class. Then, each computing node makes available this information to the map tasks that it runs, thus allowing them to determine the list and the relative position of the compressed data blocks in their HDFS data blocks.

\subsubsection{Managing Disalignments between Compressed Data Blocks and HDFS Data Blocks}
\label{subsec:disalignments}

When uploading a large compressed splittable file on HDFS, it is likely that several of its compressed data blocks would be broken into parts located on different HDFS data blocks, because of the partitioning strategy used by the distributed file system.  An example of such a case is discussed in Figure \ref{fig:layout2}. The file is initially stored as a whole on a local file system (Figure \ref{fig:layout2}(a)). If uploaded without specifying any splitting strategy, it would be partitioned into  separate parts independently of the compressed data blocks, as pictured in Figure \ref{fig:layout2}(b). This would imply a severe performance overhead when reading the content of compressed data blocks spawn across different parts.

Here, a first possible solution, denoted as {\em Compressed Block Split} strategy, would be to model as input splits all the compressed data blocks existing in a compressed file (see Figure \ref{fig:layout2}(c)). However, this strategy may imply a performance overhead because the typical size of compressed data blocks is usually orders of magnitude smaller than those of the HDFS data blocks. Thus, the number of input splits would be much larger than the number of HDFS data blocks. 


A more efficient solution, here denoted as {\em Enhanced Split} strategy, is to fit several compressed data blocks into the same Hadoop input split and, then, have each map task query a local index listing the offset of all the single compressed data blocks existing in a split (see Figure \ref{fig:layout2}(d)).  At this point, when processing compressed data blocks in a split, two cases may occur:

\begin{itemize}
\item{\bf standard case:} the compressed data block is entirely contained in a single HDFS data block. In such a circumstance, it is retrieved using the information contained in the index and, then, decompressed using the considered Codec.

\item{\bf exceptional case:} the compressed data block is physically divided by HDFS into two parts, $p_{1}$ and $p_{2}$. These parts are located on two HDFS data blocks but are assigned to the same input split. In such a case, a copy of $p_{2}$ is automatically pulled from the Datanode holding it. Then, $p_{1}$ and $p_{2}$ are properly concatenated to obtain $p$. The resulting compressed data block is decompressed using the Codec decompression function.
\end{itemize}




\subsection{The architecture of the Splittable Compressor Meta-Codec}
\label{subsec:codec}

This Meta-Codec consists of a library of abstract Java classes and interfaces implementing a standard Hadoop splittable Codec for the compression of FASTA/Q files, but without any compression/decompression routine. 

 

\label{subsubsec:customcodec}

Its architecture  is based on a specialization of the generic compressors and decompressors interface coming with Hadoop and targeting block-based Codecs. It offers the possibility to automatically assemble a compressed file as a set of compressed data blocks while maintaining their index using an explicit representation, as described in Section \ref{subsec:inferring}. In addition, the compressed data blocks are organized according to the Enhanced Split strategy (see Section \ref{subsec:disalignments}). 
Also the creation of the compressed data blocks index is automatically managed by our Meta-Codec, which also provides the ability to share the content of the index with all nodes of an Hadoop distributed system so to allow for each node to know the exact boundaries of the compressed data blocks it has to process. Additional details regarding the architecture of this Meta-Codec are  given  in Figure 1 of the \SM. Here we limit ourselves to mention that it  includes the following Java classes.


\begin{itemize}
\item{\texttt{CodecInputFormat.}}  It fetches the list of compressed data blocks existing in a compressed file and sends it to all the nodes of an Hadoop cluster together with the instructionts required for their decompression. Then, it defines the input splits as containers of compressed data blocks. These operations are compressor-dependent and require the implementation of several abstract methods like \texttt{extractMetadata}, to extract the metadata from the input file, and \texttt{getDataPosition}, to point to the starting address of the first compressed data block. 
\item{\texttt{NativeSplittableCodec.}}  Assuming the compression/decompression routines for a particular Codec are available as a standard library installed on the underlying operation system, it simplifies its integration in the Codec under development.
\item{\texttt{CodecInputStream.}} It reads the compressed data blocks existing in a HDFS data block, according to the input split strategy defined by the \texttt{CodecInputFormat}. The compressed data blocks are decompressed on-the-fly by invoking the decompression function of the considered compressor and returned to the main application. Some of these operations are compressor-dependent and require the implementation of the \texttt{setParameters} abstract method. This method is used to pass to the Codec the command-line parameters required by the compressor, e.g execution flags, in order to correctly decompress the compressed data blocks.
\item{\texttt{CodecDecompressor.}} It decompresses the compressed data blocks given by the \texttt{CodecInputStream}. It requires the implementation of the \texttt{decompress} abstract method.
\item{\texttt{NativeCodecDecompressor.}} It decompresses the compressed data blocks given by the \texttt{CodecInputStream}. It requires the implementation of the \texttt{decompress} method through the native interface.
\end{itemize}

\subsection{The architecture of the Universal Compressor Meta-Codec}

\label{subsec:UC}

This Meta-Codec is a software component able to automatically expose as a HU splittable Codec the compression/decompression routines offered by a given standard compressor.  As opposed to the  {\bf Splittable Compressor Meta-Codec}, requiring some programming, it works  as a ready-to-use black box, since the only information it needs is the set of command lines to be used for compressing and for decompressing an input file by means of a standard compressor.

Assuming there is an input file to compress in a splittable way, this method works by splitting the file into uncompressed data blocks and, then, compressing each uncompressed data block using an external compression application according to the command line given at configuration time. As for the {\bf Splittable Compressor Meta-Codec},
compressed data blocks are organized following the Enhanced Split strategy (see Section \ref{subsec:disalignments}). 

The resulting file will use an index for the explicit representation of the compressed data blocks existing therein (see Section \ref{subsec:inferring}) based on the following format.

\begin{itemize}
	\item {\bf compression\_format}: A unique id number telling the Codec format used for this file. 
	\item {\bf compressed\_data\_blocks\_number}: Number of compressed data blocks existing in the file.
	\item {\bf blocks\_sizes\_list}: List of the size of all the compressed data blocks included in the file.  
	\item {\bf uncompressed\_block\_size}: The size of the data structure used for decompressing the compressed data blocks.  
\end{itemize}

 
 The decompression is achieved by exploiting the information contained in the aforementioned index.




The usage of this Meta-Codec assumes the possibility of parking as files on a local device the content of the (un)compressed data blocks to process. For efficiency reasons, these are saved on the local RAM disk, a virtual device usable as a disk but with the same performance of memory.

The Java classes for this Meta-Codec, shown in Figure 2 of the \SM, are the following.

\begin{itemize}
    \item \texttt{Algo}. Contains the command-line instructions of a particular compressor, defined through the configuration file.
	\item \texttt{UniversalCodec}. Contains fields and methods for managing data compression and decompression.
	\item \texttt{UniversalInputFormat}. Extends the \texttt{CodecInputFormat} class, implementing the methods according to the compressed file structure.
	\item \texttt{UniversalDecompressor}. Extends the \texttt{CodecDecompressor} class, implementing the method \texttt{decompress}, according to the command-line commands of the \texttt{Algo} object.
\end{itemize}

\section{Results and Discussion}
\label{sec:experiments}

In order to quantify  the advantages of deploying  FASTA/Q Codecs in Hadoop via our methods, we perform the following experiments.



\begin{itemize}
        \item{\bf Experiment 1: An assessment  of disk space savings}. The aim here is to determine the possible disk space savings achievable thanks to the adoption of a specialized HU or HS Codec, when storing  FASTA/FASTQ files on the Hadoop HDFS distributed file system, with respect to the usage of general-purpose HS  Codecs available in Hadoop.

    \item {\bf Experiment 2: An assessment of the possible performance loss due to the usage of an HU Codec against an HS Codec}.  The aim here is to evaluate the potential performance loss that is experienced when processing a compressed file using a compressor obtained  by means of our {\bf Universal Compressor Meta-Codec } rather than using a compressor obtained via the {\bf Splittable Compressor Meta-Codec}. This experiment is implemented by comparing HU\_DSRC, obtained via the first Meta-Codec vs  HS\_DSRC, obtained via the latter Meta-Codec.


    \item{\bf Experiment 3: An assessment  of reading times savings}. The aim  here is to determine if the trade-off between the cost to be paid for reading and unpacking compressed FASTA/Q files, once compressed with an HU Codec, and the time saved thanks to the smaller amount of data to read from HDFS is positive. Following the methodology  used in \cite{fastdoop}, this experiment is implemented by benchmarking a  very simple Hadoop application. It runs only map tasks whose goal is to count the number of occurrences of the letters $\{A,C,G,T,N\}$ in the input sequences, without producing any output. That is, the application spends most of its time reading data from HDFS. 
    
    \item {\bf Experiment 4: An assessment of network communication time overhead savings}.  The aim here is to establish if the smaller amount of network traffic due to the reduced number of map tasks needed to process a  FASTA/Q file compressed with an HU Codec has a beneficial effect on the overall shuffle time of an application, compared to the case where the input file is uncompressed. This experiment is implemented by benchmarking an application where each map task counts the number of occurrences of the letters $\{A,C,G,T,N\}$, in each of the sequences read from an input file. Once finished, the map task emits, as output, the overall count for each of the considered sequences. The reduce tasks gather and aggregate the output of all map tasks, and print on output the overall number of occurrences of each distinct letter.  That is, the execution of this experiment requires a communication activity between map and reduce tasks that is proportional to the number of map tasks being used.
    
\end{itemize}




%


\subsection{Experimental Setting}

\subsubsection{Choice of Compression Codecs: Standard Specialized or Available in Hadoop}
\label{subsec:compressors}

For our experiments, all the standard splittable general-purpose compression Codecs available with Hadoop have been considered: BZIP2 \cite{Bzip2}, LZ4 \cite{Lz4} and ZSTD \cite{Zstd}.

As for the specialized  FASTA/Q files compressors, we have developed a set of compression Codecs based on SPRING \cite{spring}, DSRC \cite{roguski2014dsrc}, Fqzcomp \cite{bonfield2013compression}, MFCompress \cite{pinho2013mfcompress}.  These have been chosen, with independent experiments, as they cover the range of possibilities in terms of the trade-off compression and time. A list of all these Codecs is reported in Table \ref{tab:encoders}, with their relevant features for this research. We recall from the Introduction, for the convenience of the reader, the terminology we use for denoting these compressors: we use the prefix HS when referring to compressors that are already present in Hadoop or that have have been incorporated in it using our {\bf Splittable Compressor Meta-Codec}, and the prefix HU when the incorporation has been made with our {\bf Universal Compressor Meta-Codec }.

It is to be remarked that while the general purpose compressors have been designed to compress well and be fast in compression/decompression times, the specialized ones are not so uniform with respect to this design criteria. For instance, HU\_SPRING compresses very well, but it is very slow in compression/decompression times, while HU\_DSRC offers a good balance of those aspects. To place every compressor at a peer, we use their  default settings. 


%


\begin{table}[t]
    \centering
\begin{tabular}{lccc}
\textbf{Compressor} & \textbf{Input Format} & \textbf{Implementation} \\
 & \textbf{Type} & \\
  \hline
BZIP2 \cite{Bzip2} & Any file & HS\\
LZ4 \cite{Lz4} & Any file & HS\\
ZSTD \cite{Zstd} & Any file & HS\\
DSRC \cite{roguski2014dsrc} & FASTQ files & HS/HU \\
Fqzcomp \cite{bonfield2013compression} & FASTQ files & HU \\
MFCompress \cite{pinho2013mfcompress} & FASTA files & HU\\
SPRING \cite{spring} & FASTA/Q files & HU\\


\end{tabular}
\caption{List of splittable Codecs considered in our experiments. For each splittable Codec it is reported: 1) the originating compressor; 2) the input format it supports; 3) whether or not it has  been developed using our {\bf Splittable Compressor Meta-Codec} (HS) or our {\bf Universal Compressor Meta-Codec} (HU) or directly supported (HS). }
\label{tab:encoders}
\end{table}


\subsubsection{Datasets}
We have used for our experiments a collection of FASTQ and FASTA files, of different sizes. The FASTQ files contain a set of reads extracted from a collection of genomic sequences coming from the Pinus Taeda genome \cite{PinusTaeda2013}, while the FASTA files contain a set of reads extracted from a collection of genomic sequences coming from the Human genome \cite{Human2008}. We have chosen these datasets because they are so large to represent a relevant benchmark for the type of experiment we were interested in Section 4 of the \SM.  Moreover, the choice of using collection of reads is to consider files that are the end product of HTS technologies.

\subsubsection{Hardware}
The  testing platform used for our experiments is a $9$ nodes Linux-based Hadoop cluster, with one node acting as \textit{resource manager} and the remaining nodes being used as workers. Each node of this cluster is equipped with two 8-core Intel Xeon E3-12@2.70 GHz processor and 32GB of RAM. Moreover, each node has a 200 GB virtual disk reserved to HDFS, for an overall capacity of about 1.6 TB. All the experiments have been performed using the Hadoop 3.1.1 software distribution.


\subsection{Analysis of the experiments}

\subsubsection{Experiment 1: Specialized compression yields significant disk space savings on Hadoop.} The results of this experiment, reported in Tables \ref{tab:bs_fasta}-\ref{tab:bs_fastq}, confirm the ability of the specialized HU and HS Codecs, i.e. the ones that have been imported in Hadoop using our methods, to reach a compression ratio much higher than that of generic HS  Codecs already available in Hadoop. This is witnessed by the much smaller number of HDFS data blocks needed to store a distributed compressed representation of each file, with respect to uncompressed files. 




\begin{table}
    \centering
    \resizebox{\textwidth}{!}{
    \begin{tabular}{c|c|c|c|c|c|c}
    Dataset & NoCompress & HS\_BZIP2 & HS\_LZ4 & HS\_ZSTD & HU\_SPRING & HU\_MFCompress \\
     \hline
    16G & 128 & 24 & 58 & 31 & 18 & 19 \\
    32G & 256 & 47 & 116 & 62 & 35 & 38 \\
    64G & 512 & 94 & 231 & 124 & 69 & 76 \\
    96G & 768 & 141 & 346 & 185 & 104 & 113 \\
    \end{tabular}
    }
    \caption{Size of the FASTA input datasets, in terms of HDFS data blocks, when compressed with general-purpose and FASTA specialized compression Codecs. The size of each HDFS data block is 128 MB.}
    \label{tab:bs_fasta}
\end{table}

\begin{table}
    \centering
    \resizebox{\textwidth}{!}{
    \begin{tabular}{c|c|c|c|c|c|c|c|c}
    Dataset & NoCompress & HU\_DSRC & HS\_DSRC & HU\_Fqzcomp & HS\_BZIP2 & HS\_LZ4 & HS\_ZSTD & HU\_SPRING \\
     \hline
    16G & 128 & 22 & 20 & 20 & 25 & 60 & 33 & 20 \\
    32G & 256 & 44 & 40 & 40 & 49 & 119 & 64 & 40 \\
    64G & 512 & 90 & 80 & 82 & 99 & 241 & 130 & 81 \\
    96G & 768 & 122 & 122 & 110 & 150 & 364 & 196 & 103 \\
     \end{tabular}
    }
    \caption{Size of the FASTQ input datasets, in terms of HDFS data blocks, when compressed with general-purpose and FASTQ specialized compression Codecs. The size of each HDFS data block is 128 MB.}
    \label{tab:bs_fastq}
\end{table}

\subsubsection{Experiment 2: The performance overhead of our {\bf Universal Compressor Meta-Codec} with respect to our {\bf Splittable Compressor Meta-Codec} is negligible.}

\begin{figure}
    \centering
    \includegraphics[scale=.4]{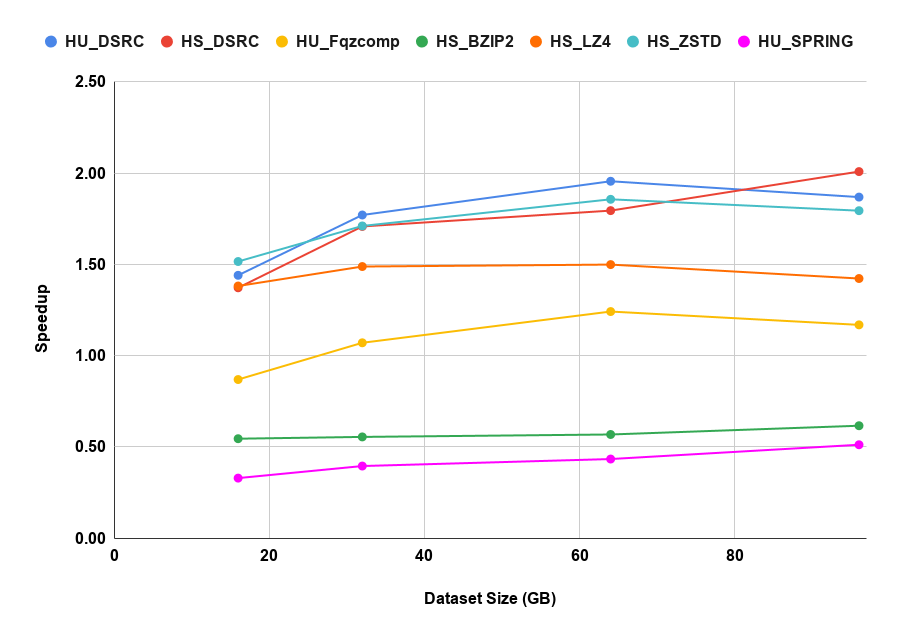}
    \caption{Execution time speedup measured while running the first benchmarking task when considering FASTQ compressed datasets of increasing size and different compressors, with respect to the execution on the equivalent uncompressed datasets.}
    \label{fig:task1FQGARR}
\end{figure}

The decompression time performance guaranteed by our {\bf Universal Compressor Meta-Codec } when executing a particular compressor is very similar to that of a specialized implementation of the same compressor by means of our {\bf Splittable Compressor Meta-Codec}. This is clearly visible in Figures \ref{fig:task1FQGARR} and \ref{fig:task2FQGARR}, where we report the performance of HS\_DSRC and HU\_DSRC. Indeed, the two Codecs exhibit very similar performance, but the one based on our {\bf Universal Compressor Meta-Codec } took few minutes to be developed while the specialized one required non trivial programming skills as well as several days of work.

\subsubsection{Experiment 3: a careful use of compression yields significant reading-times savings on Hadoop.}

Space savings may turn into I/O time slow-down, when the decompression procedure is slow. Such a trade-off is well known for generic standard compressors. Here we study it in regard to HU specialized Codecs. Indeed, such a trade-off  is clearly visible when comparing, e.g., the performance of HU\_DSRC with those of HU\_SPRING. As reported in Tables \ref{tab:bs_fasta} and \ref{tab:bs_fastq}, FASTQ files compressed with HU\_SPRING require about a smaller number of HDFS data blocks to process than those compressed with HU\_DSRC. Despite this, the performance of HU\_DSRC when used in the first benchmarking task are much better than that of HU\_SPRING because of its much faster decompression routines. 

In details, the best performance is achieved by HS\_DSRC, HU\_DSRC and HS\_ZSTD, but for different reasons: the first two because of their more efficient compression algorithm, the third because of its faster decompression routines. We also observe that the speed-up achieved by HS\_DSRC increases with the input file size. To explain this, consider that when managing the 16G input file, HS\_DSRC returns a number of HDFS data blocks to process that is smaller than the number of available processing cores. So, not all the available processing capability of the cluster is exploited. When the size of the input increases to 32G, the number of HDFS data blocks gets larger and allows to use all the available processing cores, thus resulting in an improved overall efficiency. This speed-up gets increasing because, as well as the input size grows, the number of HDFS data blocks to process per core increases as well, giving Hadoop the possibility to reschedule tasks over the cores having a smaller workload.



On the bottom side, the HS\_BZIP2 and HU\_SPRING are the ones exhibiting the worst performance, because of their very slow decompression routines.

\subsubsection{Experiment 4: a careful use of compression may yield significant network-overhead savings on Hadoop.}

\begin{figure}
    \centering
    \includegraphics[scale=.4]{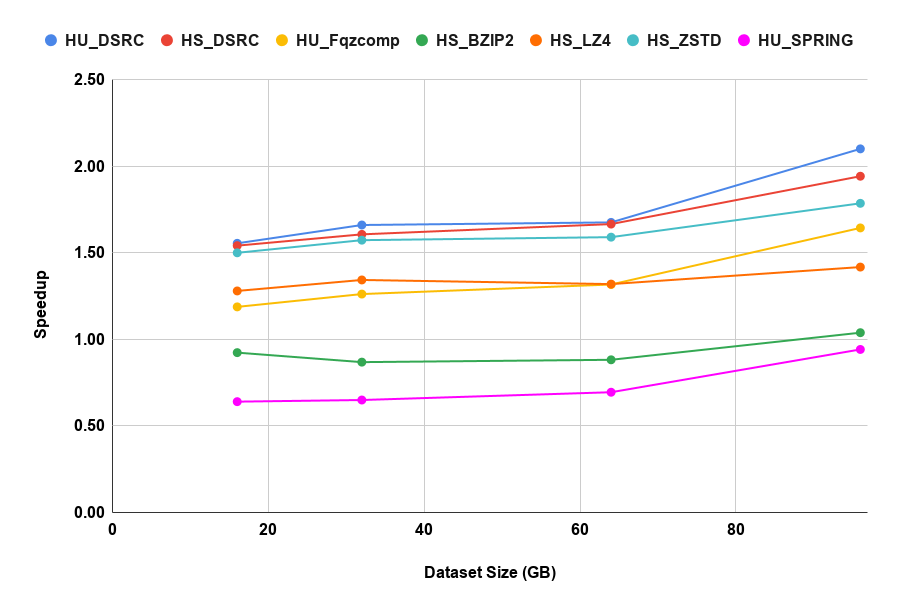}
    \caption{Execution time speedup measured  while running the second benchmarking task on FASTQ format data when considering compressed datasets of increasing size and different compressors, with respect to the execution on the equivalent uncompressed datasets.}
    \label{fig:task2FQGARR}
\end{figure}

The smaller amount of HDFS data blocks required to store a compressed file yields a beneficial effect also on the network overhead required by Hadoop to recombine the output of the map tasks and, consequently, on the overall execution time. This is visible in Figures \ref{fig:task2FQGARR} and \ref{fig:task2FAGARR}, where we observe that the usage of compression allows for a significant speed-up, even when running more complex applications that the one considered in experiment 3.

\begin{figure}
    \centering
    \includegraphics[scale=.4]{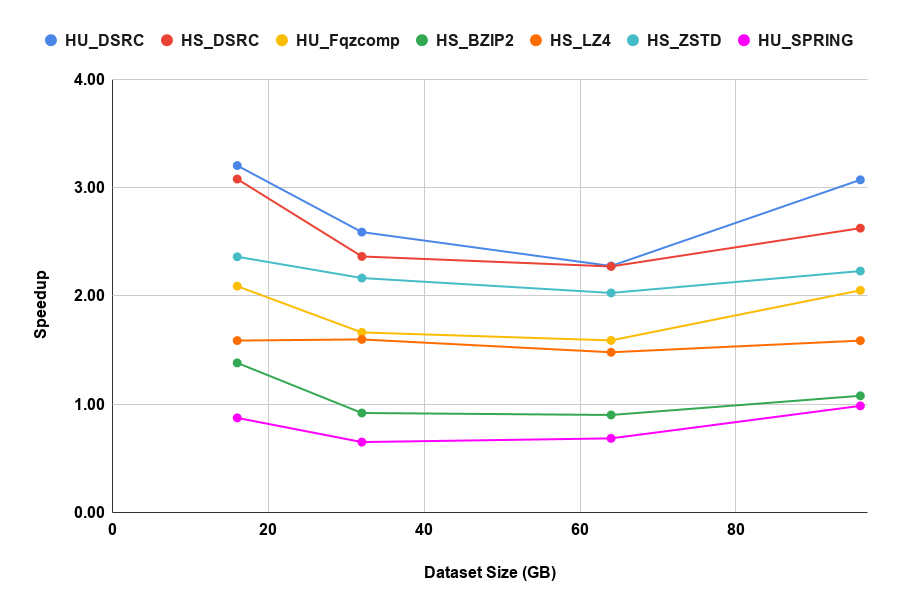}
    \caption{Time spent running the shuffle phase during the second benchmarking task when considering compressed FASTQ-format datasets of increasing size and different compressors, compared to the execution of the same task on uncompressed datasets.}
    \label{fig:shuffleFQGARR}
\end{figure}

Interestingly, here the benchmarking task run using  HU\_DSRC and HS\_DSRC is faster than the one run using HS\_ZSTD  (see Figure \ref{fig:task2FQGARR}). The reason is that the smaller number of compressed data blocks produced by the DSRC algorithm implies a smaller number of Hadoop map tasks to be concurrently run for analyzing the input dataset thus reducing, in turn, the network overhead required for feeding the reduce tasks with the output of the map tasks. The smaller network overhead achievable by using either HS\_DSRC or HU\_DSRC rather than HS\_ZSTD is witnessed by the reduced shuffle time, as observable in Figure \ref{fig:shuffleFQGARR}.

\begin{figure}
    \centering
    \includegraphics[scale=.4]{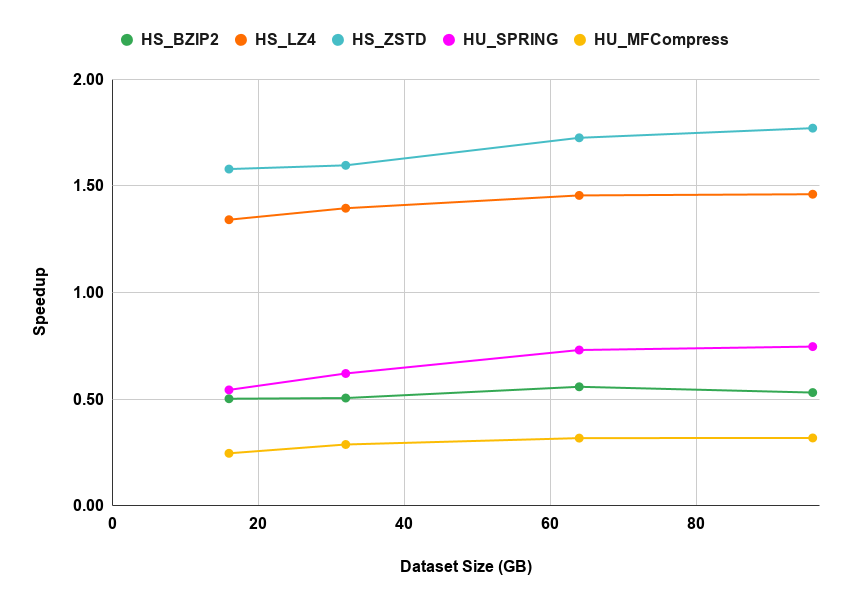}
    \caption{Execution time speedup measured while running the first benchmarking task when considering FASTA compressed datasets of increasing size and different compressors, with respect to the execution on the equivalent uncompressed datasets.}
    \label{fig:task1FAGARR}
\end{figure}

\begin{figure}
    \centering
    \includegraphics[scale=.4]{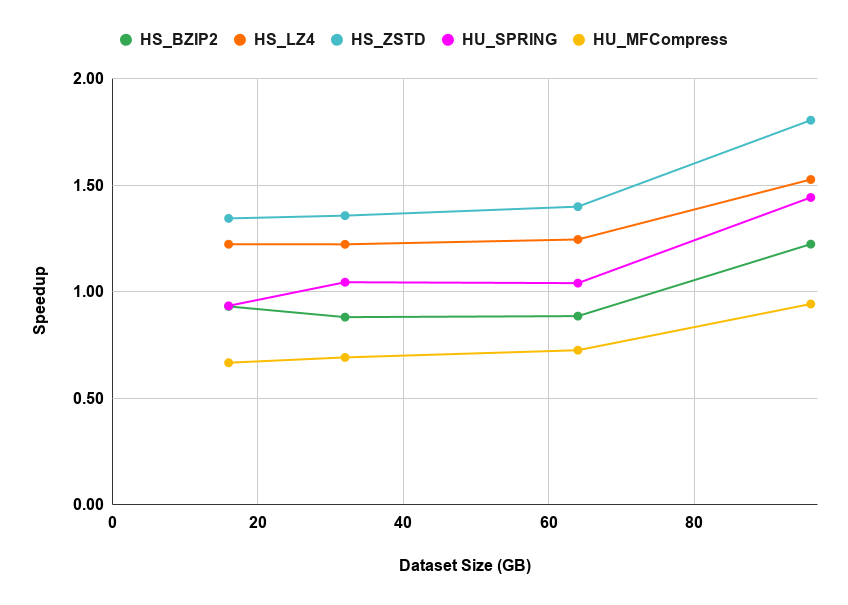}
    \caption{Execution time speedup measured  while running the second benchmarking task on FASTA format data when considering compressed datasets of increasing size and different compressors, with respect to the execution on the equivalent uncompressed datasets.}
    \label{fig:task2FAGARR}
\end{figure}

\begin{figure}
    \centering
    \includegraphics[scale=.4]{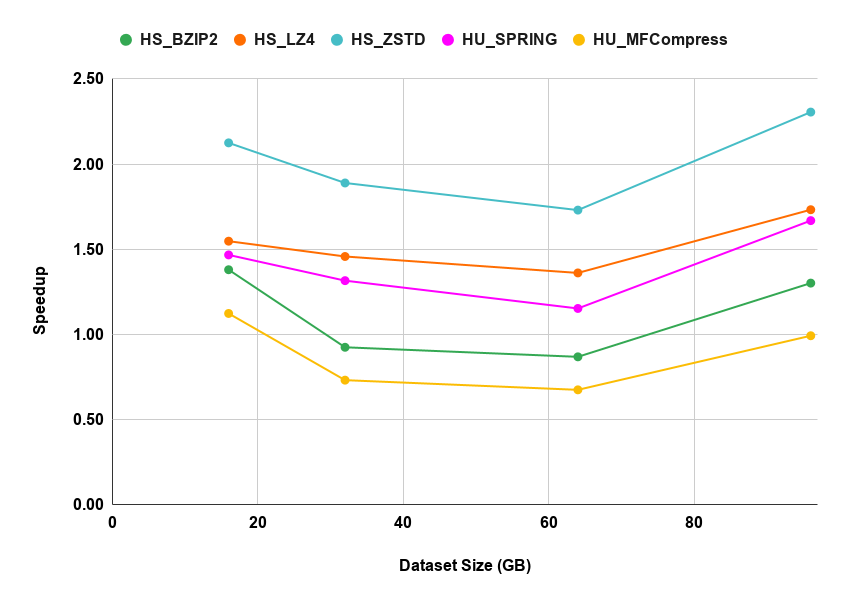}
    \caption{Time spent running the shuffle phase during the second benchmarking task when considering compressed FASTA-format datasets of increasing size and different compressors, compared to the execution of the same task on uncompressed datasets.}
    \label{fig:shuffleFAGARR}
\end{figure}

\section{Conclusions}
We have provided two general methods that can be used to transform standard FASTA/Q data compression programs into Hadoop splittable data compression Codecs. Being the methods general, they can be used for specialized standard compression programs that will be developed in the future.  Another main characteristic of our methods is that they require very little,  or none at all, programming and knowledge of Hadoop to carry out a rather complex task. Our methods apply also to the Apache Spark framework, when used to process FASTA/Q files stored on the Hadoop File System.

We have also shown that the use of specialized FASTA/Q Hadoop Codecs, not available before this work,  is advantageous in terms of space and time savings. That is, we provide effective and readily usable tools that have a non-negligible effect on saving costos in genomic data storage and processing within Big Data Technologies. 

\section*{Acknowledgements}
All authors would like to thank the computing time on a cutting edge OpenStack Virtual Datacenter for this research made available by GARR. Discussions with Simona Ester Rombo in the early stages of this research have been helpful.

\section*{Funding}
G.C., R.G. and U.F.P. are partially supported by GNCS Project 2019 \vir{Innovative methods for the solution of medical and biological big data}. R.G. is additionally supported by  MIUR-PRIN project \vir{Multicriteria Data Structures and Algorithms: from compressed to learned indexes, and beyond} grant n. 2017WR7SHH. 
U.F.P. and F.P. are partially supported by Universit\`{a} di Roma - La Sapienza Research Project 2018 \vir{Analisi, sviluppo e sperimentazione di algoritmi praticamente efficienti}.

\bibliographystyle{abbrv}
\bibliography{biblio}

\end{document}


\maketitle

\begin{abstract}
Additional details about the \MM are provided in this document.
\end{abstract}

\section{The \MR Programming Paradigm and Hadoop}
\label{sec:mr-hadoop}

	\subsection{The Paradigm} \label{sec:mr}

	\MR \cite{dean2008mapreduce} is a paradigm for the processing of large amounts of data on a distributed computing infrastructure. Assuming the input data is organized as a set of \KV{key}{value} pairs, it is based on the definition of two functions. The {\em map} function processes an input \KV{key}{value} pair and returns a (possibly empty) intermediate set of \KV{key}{value} pairs. The {\em reduce} function merges all the intermediate values sharing the same \empty{key} to form a (possibly smaller) set of values. These functions are run, as tasks, on the nodes of a distributed computing framework. All the activities related to the management of the lifecycle of these tasks as well as the collection of the map function results and their transmission to the reduce functions are transparently handled by the underlying framework (\emph{implicit parallelism}), with no burden on the programmer.

	\subsection{Apache Hadoop} \label{sec:hadoop}
	
	Apache Hadoop is the most popular framework supporting the \MR paradigm. It allows for the execution of distributed computations thanks to the interplay of two architectural components: YARN (\emph{Yet Another Resource Negotiator}) \cite{vavilapalli2013apache} and HDFS (\emph{Hadoop Distributed File System}) \cite{HDFS}. YARN manages the lifecycle of a distributed application by keeping track of the resources available on a computing cluster and allocating them for the execution of application tasks modeled after one of the supported computing paradigms. HDFS is a distributed and block-structured file-system designed to run on commodity hardware and able to provide fault tolerance through replication of data. 
	
	A basic Hadoop cluster is composed  of a single \emph{master node} and multiple \emph{worker nodes}. The master node arbitrates the assignment of computational resources to applications to be run on the cluster and  maintains an index of all the directories and the files stored in the HDFS distributed file system. Moreover, it  tracks the worker nodes physically storing the HDFS data blocks making up these files. The worker nodes host a set of \emph{worker}s (also called \emph{Containers}), in charge of running the map and reduce tasks of a \MR application, as well as using the local storage to maintain a subset of the HDFS data blocks.
	
	One of the main characteristics of Hadoop is its ability to exploit \emph{data-local} computing. By this term, we mean the possibility  to move applications closer to the data (rather than the vice-versa). This allows to greatly reduce network congestion and increase the overall throughput of the system when processing large amounts of data. Moreover, in order to reliably maintain files and to properly balance the load between different nodes of a cluster, large files are automatically split into smaller HDFS data blocks, replicated and spread across different nodes.

\section{Specialized Compressors Supported by means of our {\bf Splittable Compressor Meta-Codec}}
\label{sec:DSRC}

Among the many compression algorithms specialized for genomic data \cite{Numanagic2016}, DSRC is the only featuring a splittable Codec among the data compression tools achieving the best performance, based on benchmarking, when dealing with FASTA/Q files. 
 
It represents a robust testbed for our solution because its original implementation has been developed in C++ and its integration within a Java Codec is not trivial to realize. 
 
A DSRC standard compressed file is organized in three parts.
 
 \begin{itemize}
 	\item {\bf Body.} It contains a set of compressed data blocks. Each of these is compressed and can be decompressed independently from the others. The default size of each compressed data block is 10MB.
 	\item {\bf Header.} It reports the number of compressed data blocks existing in that file, the size of the footer and its relative position inside the file.
 	\item {\bf Footer.} It reports the size of each compressed data block and the flags used for its compression.
 \end{itemize} 
 


\subsection{Implementation details}
\label{subsec:specialpurpose}
The special-purpose Codec supporting DSRC, HS\_DSRC, has been obtained following our {\bf Splittable Compressor Meta-Codec}, as described in Section 2.3 of the \MM. It required the development of two Java classes:  \texttt{DSRCInputFormat} and \texttt{DSRCCodec}. In particular, \texttt{DSRCCodec} uses the JNI framework\cite{Jni} to load in memory and instantiate the dynamic library containing the DSRC native implementation. 
Then, it uses the \texttt{DSRCInputFormat} class to extract the information regarding the DSRC parameters and the list of compressed data blocks, according to the DSRC format. In addition, this class initializes the \texttt{CodecInputStream} object, pointing to the file to be decompressed during the execution of a job.  Finally, it runs the  \texttt{NativeCodecDecompressor decompress} method on each compressed data block to obtain its decompressed version.

\section{Specialized Compressors Supported by means of our {\bf Universal Compressor Meta-Codec}}

In this Section we provide details about the work done for incorporating in Hadoop the specialized compressors reported in Section 3.1.1 of the \MM, using our {\bf Universal Compressor Meta-Codec}.

For each compressor, the only step required to support it is the definition of a set of properties stating the supported input file types and the command-line required for compressing and decompressing a generic input file. Let X be the unique name denoting the compressor to be supported and F the file being processed, the following command line properties are available for its integration:

\begin{itemize}
	\item{\texttt{uc.X.compress.cmd}}: the command line to be used for compressing F using X.
	\item{\texttt{uc.X.decompress.cmd}}: the command line to be used for decompressing F using X.
	\item{\texttt{uc.X.io.input.flag}}: the command line flag used to specify the input filename.
	\item{\texttt{uc.X.io.output.flag}}: the command line flag used to specify the output filename.
	\item{\texttt{uc.X.compress.ext}}: the extension used by X for saving a compressed copy of F.
	\item{\texttt{uc.X.decompress.ext}}: the extension used by X for saving a decompressed copy of X ("fastq" by default).
	\item{\texttt{uc.X.io.reverse}}: if X requires the output file name to be specified before the input file name, it is set to \emph{true}. {false}, otherwise. 
\end{itemize}

In Table \ref{tab:my_label}, the command lines used for integrating the target specialized compressors using our {\bf Universal Compressor Meta-Codec} are reported.

\begin{table}
    \centering
    \resizebox{\columnwidth}{!}{
    \begin{tabular}{|l|l|l|l|l|l|}
        \hline
        \multicolumn{1}{|c|}{Properties} & \multicolumn{5}{c|}{Compressors}\\\cline{2-6}
         & SPRING (for FASTQ) & SPRING (for FASTA) & DSRC & FqzComp & MFCompress\\
        \hline
        uc.X.compress.cmd & spring -c & spring -c --fasta-input & dsrc c -t8 & fqz\_comp & MFCompressC -t 8 -p 8\\
        uc.X.decompress.cmd & spring -d & spring -d & dsrc d -t8 & fqz\_comp -d & MFCompressD -t 8\\
        uc.X.io.input.flag & -i & -i &  & & \\
        uc.X.io.output.flag & -o & -o &  & & -o \\
        uc.X.compress.ext & .spring & .spring & .dsrc & .fqz & .mfc \\
        uc.X.decompress.ext &  & .fasta &  &  & .fasta \\
        uc.X.io.reverse &  &  &  &  & true \\
        \hline
    \end{tabular}
    }
    \caption{Command line properties required for supporting several specialized compressors using our {\bf Universal Compressor Meta-Codec}}
    \label{tab:my_label}
\end{table}

\section{Datasets}
\label{sec:datasets}

The FASTQ files used in our experiments contain a set of reads extracted uniformly at random from a collection of genomic sequences coming from the Pinus Taeda genome \cite{PinusTaeda2013}. The FASTA files used in our experiments contain a set of reads extracted uniformly at random from a collection of genomic sequences coming from the Human genome \cite{Human2008}.
Details about the files included in these datasets are reported in Table \ref{tab:FAdataset} and Table \ref{tab:FQdataset}.

\begin{table}[!ht]
    \centering
    \begin{tabular}{|l|r|c|}
        \hline
        Name & \# of reads & Avg. read length \\
        \hline
        16GB & 96,407,378 & 100 \\
        32GB & 192,653,438 & 100 \\
        64GB & 385,306,876 & 100 \\
        96GB & 577,960,314 & 100 \\
        \hline
    \end{tabular}
    \caption{Files included in the FASTA dataset used in our experiments}
    \label{tab:FAdataset}
\end{table}
\begin{table}[!ht]
    \centering
    \begin{tabular}{|l|r|c|}
        \hline
        Name & \# of reads & Avg. read length \\
        \hline
        16GB & 44,681,859 & 151 \\
        32GB & 89,363,718 & 151 \\
        64GB & 178,727,437 & 151 \\
        96GB & 268,091,154 & 151 \\
        \hline
    \end{tabular}
    \caption{Files included in the FASTQ dataset used in our experiments}
    \label{tab:FQdataset}
\end{table}

\begin{figure}[!ht]
    \centering
    \includegraphics[scale=.45]{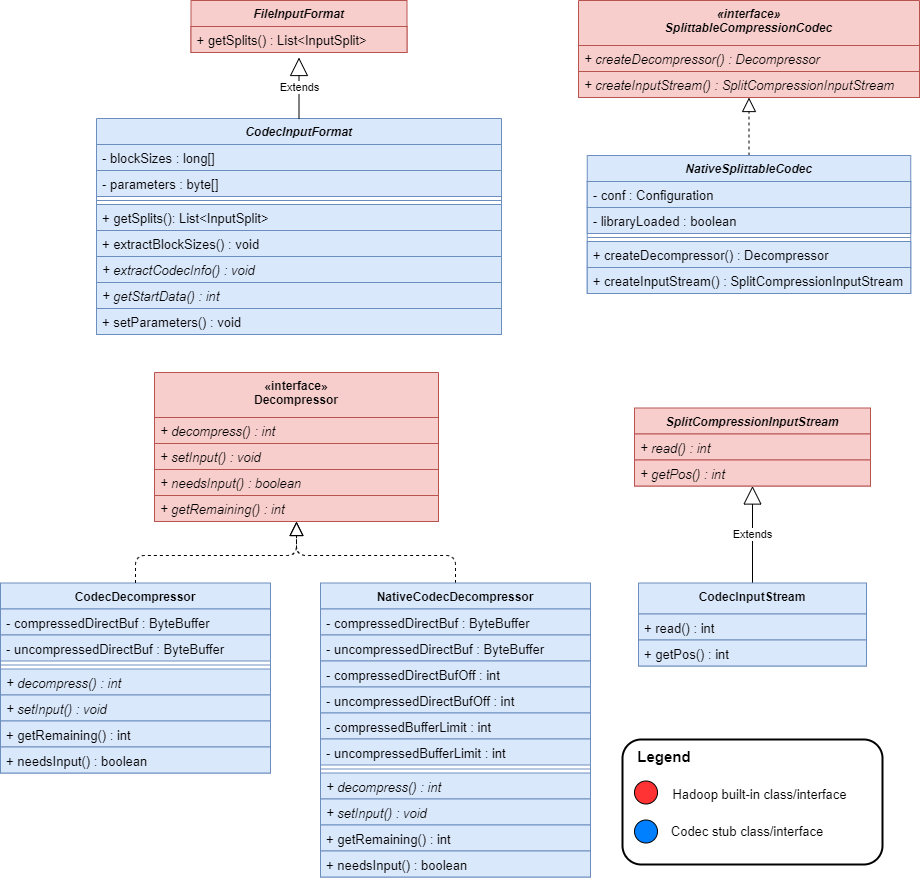}
    \caption{UML class diagram of our \textbf{Splittable Compressor Meta-Codec}}
        \label{fig:classGeneric}
\end{figure}

\begin{figure}[!ht]
    \centering
    \includegraphics[scale=.35]{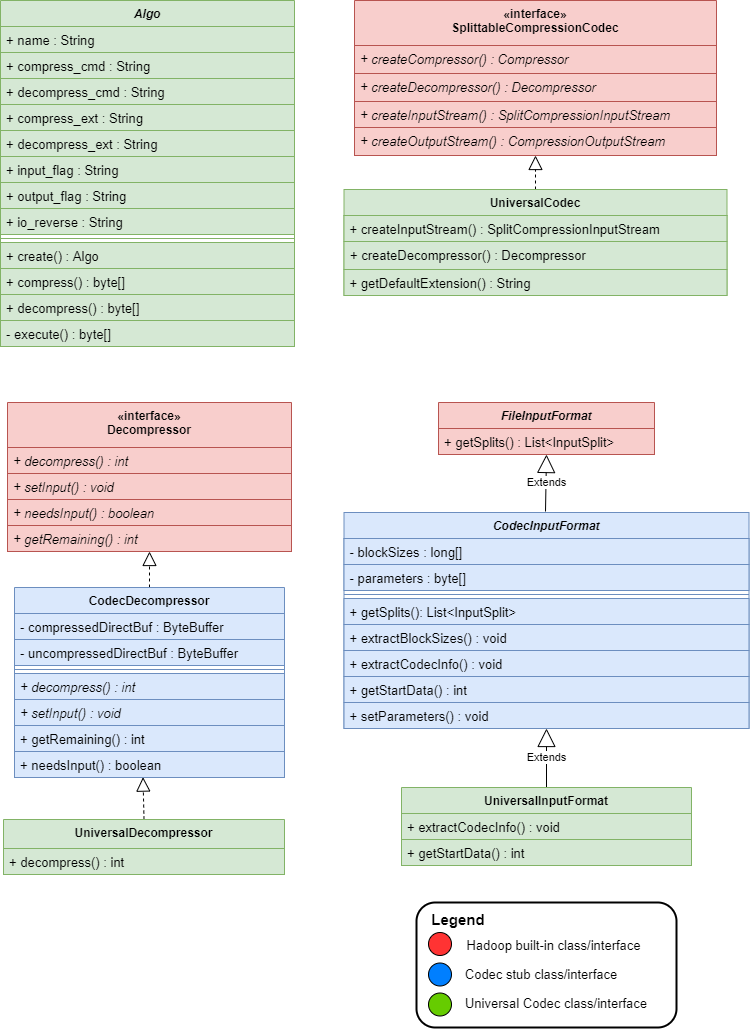}
    \caption{UML class diagram of our \textbf{Universal Compressor Meta-Codec}}
        \label{fig:classUniversal}
\end{figure}

\newpage

\begin{figure}[!ht]
    \centering
    \includegraphics[scale=.45]{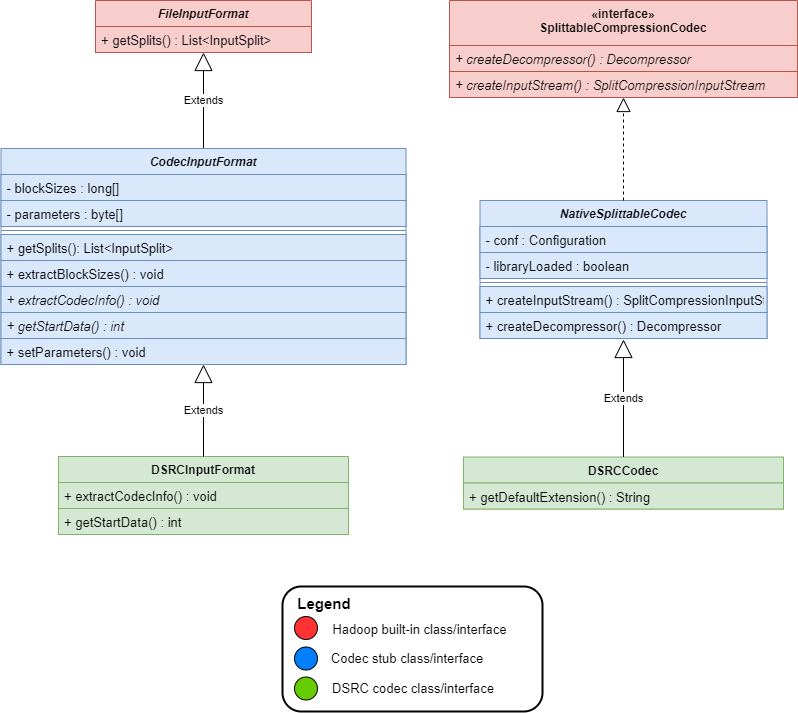}
    \caption{UML class diagram of HS\_DSRC }
        \label{fig:classDSRC}
\end{figure}

\clearpage

\bibliographystyle{abbrv}
\bibliography{biblio}